\documentclass[aps,prd,twocolumn,preprintnumbers,superscriptaddress,letterpaper,floatfix,nofootinbib,showpacs]{revtex4}
\usepackage{afterpage,epsfig,units,color,ulem,amsmath,multirow}
\begin{document}
% \pagewiselinenumbers
\preprint{FERMILAB-PUB-10-526-E}
\preprint{BNL-94523-2010-JA}
\preprint{hep-ex/1012.3391}

\newcommand{\esurf}{E$_{\mu}$}
\newcommand{\esurfspace}{E$_{\mu}$ }
\newcommand{\esurfmath}{\textrm{E}_{\mu}}

\newcommand{\ecossurf}{E$_{\mu}$cos$\theta$*}
\newcommand{\ecossurfspace}{E$_{\mu}$cos$\theta$* }
\newcommand{\ecossurfmath}{\textrm{E}_{\mu} \textrm{cos} \theta\textrm{*}}
\newcommand{\epsilonmath}{\epsilon}
\newcommand{\mwe}{mwe}
  
\newcommand{\oldcostheta}{cos$\theta$}
\newcommand{\oldcosthetaspace}{cos$\theta$ }

\newcommand{\costheta}{cos$\theta$*}
\newcommand{\costhetaspace}{cos$\theta$* }
\newcommand{\costhetamath}{\textrm{cos} \theta \textrm{*} }

\newcommand{\rpm} {N_{\mu^+}/N_{\mu^-}}
\newcommand{\ecc}{E_\mu \cos \theta\textrm{*}}


\title{Measurement of the underground atmospheric muon charge ratio using the MINOS Near Detector}



%\input{minos-authors-nim}

%\author{Jeffrey de Jong}
%\email[]{dejong@iit.edu}
%\affiliation{Illinois Institute of Technology,Chicago IL 60616}

\newcommand{\Berkeley}{Lawrence Berkeley National Laboratory, Berkeley, California, 94720 USA}
\newcommand{\Cambridge}{Cavendish Laboratory, University of Cambridge, Madingley Road, Cambridge CB3 0HE, United Kingdom}
\newcommand{\FNAL}{Fermi National Accelerator Laboratory, Batavia, Illinois 60510, USA}
\newcommand{\RAL}{Rutherford Appleton Laboratory, Science and Technologies Facilities Council, OX11 0QX, United Kingdom}
\newcommand{\UCL}{Department of Physics and Astronomy, University College London, Gower Street, London WC1E 6BT, United Kingdom}
\newcommand{\Caltech}{Lauritsen Laboratory, California Institute of Technology, Pasadena, California 91125, USA}
\newcommand{\Alabama}{Department of Physics and Astronomy, University of Alabama, Tuscaloosa, Alabama 35487, USA}
\newcommand{\ANL}{Argonne National Laboratory, Argonne, Illinois 60439, USA}
\newcommand{\Athens}{Department of Physics, University of Athens, GR-15771 Athens, Greece}
\newcommand{\NTUAthens}{Department of Physics, National Tech. University of Athens, GR-15780 Athens, Greece}
\newcommand{\Benedictine}{Physics Department, Benedictine University, Lisle, Illinois 60532, USA}
\newcommand{\BNL}{Brookhaven National Laboratory, Upton, New York 11973, USA}
\newcommand{\CdF}{APC -- Universit\'{e} Paris 7 Denis Diderot, 10, rue Alice Domon et L\'{e}onie Duquet, F-75205 Paris Cedex 13, France}
\newcommand{\Cleveland}{Cleveland Clinic, Cleveland, Ohio 44195, USA}
\newcommand{\Delhi}{Department of Physics \& Astrophysics, University of Delhi, Delhi 110007, India}
\newcommand{\GEHealth}{GE Healthcare, Florence South Carolina 29501, USA}
\newcommand{\Harvard}{Department of Physics, Harvard University, Cambridge, Massachusetts 02138, USA}
\newcommand{\HolyCross}{Holy Cross College, Notre Dame, Indiana 46556, USA}
\newcommand{\IIT}{Physics Division, Illinois Institute of Technology, Chicago, Illinois 60616, USA}
\newcommand{\Iowa}{Department of Physics and Astronomy, Iowa State University, Ames, Iowa 50011 USA}
\newcommand{\Indiana}{Indiana University, Bloomington, Indiana 47405, USA}
\newcommand{\ITEP}{High Energy Experimental Physics Department, ITEP, B. Cheremushkinskaya, 25, 117218 Moscow, Russia}
\newcommand{\JMU}{Physics Department, James Madison University, Harrisonburg, Virginia 22807, USA}
\newcommand{\LASL}{Nuclear Nonproliferation Division, Threat Reduction Directorate, Los Alamos National Laboratory, Los Alamos, New Mexico 87545, USA}
\newcommand{\Lebedev}{Nuclear Physics Department, Lebedev Physical Institute, Leninsky Prospect 53, 119991 Moscow, Russia}
\newcommand{\LLL}{Lawrence Livermore National Laboratory, Livermore, California 94550, USA}
\newcommand{\LosAlamos}{Los Alamos National Laboratory, Los Alamos, New Mexico 87545, USA}
\newcommand{\MIT}{Lincoln Laboratory, Massachusetts Institute of Technology, Lexington, Massachusetts 02420, USA}
\newcommand{\Minnesota}{University of Minnesota, Minneapolis, Minnesota 55455, USA}
\newcommand{\Crookston}{Math, Science and Technology Department, University of Minnesota -- Crookston, Crookston, Minnesota 56716, USA}
\newcommand{\Duluth}{Department of Physics, University of Minnesota -- Duluth, Duluth, Minnesota 55812, USA}
\newcommand{\Ohio}{Center for Cosmology and Astro Particle Physics, Ohio State University, Columbus, Ohio 43210 USA}
\newcommand{\Otterbein}{Otterbein College, Westerville, Ohio 43081, USA}
\newcommand{\Oxford}{Subdepartment of Particle Physics, University of Oxford, Oxford OX1 3RH, United Kingdom}
\newcommand{\PennState}{Department of Physics, Pennsylvania State University, State College, Pennsylvania 16802, USA}
\newcommand{\PennU}{Department of Physics and Astronomy, University of Pennsylvania, Philadelphia, Pennsylvania 19104, USA}
\newcommand{\Pittsburgh}{Department of Physics and Astronomy, University of Pittsburgh, Pittsburgh, Pennsylvania 15260, USA}
\newcommand{\IHEP}{Institute for High Energy Physics, Protvino, Moscow Region RU-140284, Russia}
\newcommand{\Rochester}{Department of Physics and Astronomy, University of Rochester, New York 14627 USA}
\newcommand{\RoyalH}{Physics Department, Royal Holloway, University of London, Egham, Surrey, TW20 0EX, United Kingdom}
\newcommand{\Carolina}{Department of Physics and Astronomy, University of South Carolina, Columbia, South Carolina 29208, USA}
\newcommand{\SLAC}{Stanford Linear Accelerator Center, Stanford, California 94309, USA}
\newcommand{\Stanford}{Department of Physics, Stanford University, Stanford, California 94305, USA}
\newcommand{\StJohnFisher}{Physics Department, St. John Fisher College, Rochester, New York 14618 USA}
\newcommand{\Sussex}{Department of Physics and Astronomy, University of Sussex, Falmer, Brighton BN1 9QH, United Kingdom}
\newcommand{\TexasAM}{Physics Department, Texas A\&M University, College Station, Texas 77843, USA}
\newcommand{\Texas}{Department of Physics, University of Texas at Austin, 1 University Station C1600, Austin, Texas 78712, USA}
\newcommand{\TechX}{Tech-X Corporation, Boulder, Colorado 80303, USA}
\newcommand{\Tufts}{Physics Department, Tufts University, Medford, Massachusetts 02155, USA}
\newcommand{\UNICAMP}{Universidade Estadual de Campinas, IFGW-UNICAMP, CP 6165, 13083-970, Campinas, SP, Brazil}
\newcommand{\UFG}{Instituto de F\'{i}sica, Universidade Federal de Goi\'{a}s, CP 131, 74001-970, Goi\^{a}nia, GO, Brazil}
\newcommand{\USP}{Instituto de F\'{i}sica, Universidade de S\~{a}o Paulo,  CP 66318, 05315-970, S\~{a}o Paulo, SP, Brazil}
\newcommand{\Warsaw}{Department of Physics, Warsaw University, Ho\.{z}a 69, PL-00-681 Warsaw, Poland}
\newcommand{\Washington}{Physics Department, Western Washington University, Bellingham, Washington 98225, USA}
\newcommand{\WandM}{Department of Physics, College of William \& Mary, Williamsburg, Virginia 23187, USA}
\newcommand{\Wisconsin}{Physics Department, University of Wisconsin, Madison, Wisconsin 53706, USA}
\newcommand{\deceased}{Deceased.}

\affiliation{\ANL}
\affiliation{\Athens}
\affiliation{\Benedictine}
\affiliation{\BNL}
\affiliation{\Caltech}
\affiliation{\Cambridge}
\affiliation{\UNICAMP}
%\affiliation{\CdF}
\affiliation{\FNAL}
\affiliation{\UFG}
\affiliation{\Harvard}
\affiliation{\HolyCross}
\affiliation{\IIT}
\affiliation{\Indiana}
\affiliation{\Iowa}
%\affiliation{\IHEP}
%\affiliation{\ITEP}
%\affiliation{\JMU}
\affiliation{\Lebedev}
\affiliation{\LLL}
\affiliation{\UCL}
\affiliation{\Minnesota}
\affiliation{\Duluth}
\affiliation{\Otterbein}
\affiliation{\Oxford}
\affiliation{\Pittsburgh}
\affiliation{\RAL}
\affiliation{\USP}
\affiliation{\Carolina}
\affiliation{\Stanford}
\affiliation{\Sussex}
\affiliation{\TexasAM}
\affiliation{\Texas}
\affiliation{\Tufts}
\affiliation{\Warsaw}
\affiliation{\Washington}
\affiliation{\WandM}
%\affiliation{\Wisconsin}

\author{P.~Adamson}
\affiliation{\FNAL}
%\affiliation{\UCL}
%\affiliation{\Sussex}

\author{C.~Andreopoulos}
\affiliation{\RAL}
%\affiliation{\Athens}

%\author{K.~E.~Arms}
%\affiliation{\Minnesota}

%\author{R.~Armstrong}
%\affiliation{\Indiana}

\author{D.~J.~Auty}
\affiliation{\Sussex}

%\author{S.~Avvakumov}
%\affiliation{\Stanford}

\author{D.~S.~Ayres}
\affiliation{\ANL}

\author{C.~Backhouse}
\affiliation{\Oxford}

%\author{B.~Baller}
%\affiliation{\FNAL}

%\author{B.~Barish}
%\affiliation{\Caltech}

%\author{P.~D.~Barnes~Jr.}
%\affiliation{\LLL}

\author{G.~Barr}
\affiliation{\Oxford}

\author{W.~L.~Barrett}
\affiliation{\Washington}

%\author{E.~Beall}
%\altaffiliation[Now at\ ]{\Cleveland .}
%\affiliation{\ANL}
%\affiliation{\Minnesota}

%\author{B.~R.~Becker}
%\affiliation{\Minnesota}

%\author{A.~Belias}
%\affiliation{\RAL}

%\author{R.~H.~Bernstein}
%\affiliation{\FNAL}

%\author{M.~Betancourt}
%\affiliation{\Minnesota}

%\author{D.~Bhattacharya}
%\affiliation{\Pittsburgh}

\author{P.~Bhattarai}
\affiliation{\Duluth}

\author{M.~Bishai}
\affiliation{\BNL}

\author{A.~Blake}
\affiliation{\Cambridge}

%\author{B.~Bock}
%\affiliation{\Duluth}

% ...

\collaboration{The MINOS Collaboration}
\noaffiliation

\date{\today}

%\maketitle
%

%\thanks{}

%\collaboration{MINOS}

\date{\today}

\begin{abstract}
The magnetized MINOS Near Detector, at a depth of 225 meters of water equivalent (mwe), is used to measure the atmospheric muon charge ratio.  The ratio of observed positive to negative atmospheric muon rates, using \unit[301]{days} of data, is measured to be $1.266\pm0.001(\textrm{stat.})^{+0.015}_{-0.014}(\textrm{syst.})$.  This measurement is consistent with previous results from other shallow underground detectors, and is $0.108\pm0.019(\textrm{stat.~+~syst.})$ lower than the measurement at the functionally identical MINOS Far Detector at a depth of \unit[2070]{mwe}.  This increase in charge ratio as a function of depth is consistent with an increase in the fraction of muons arising from kaon decay for increasing muon surface energies. \\
\end{abstract}
% significantly lower than that observed at the deeper detectors
%\pacs{13.85.Tp. 95.55.Vj, 95.85.Rj, 96.50.S-,98.70.Vc}

\maketitle

\section{Introduction}
\label{sec:intro}

High energy cosmic-ray primaries interact with nuclei in the upper atmosphere and produce showers which contain pions and kaons.  These secondary mesons can either interact in further collisions in the atmosphere, or decay to produce atmospheric muons.  Since the majority of primary cosmic-rays are protons, there is an excess of positively charged mesons in the showers, and consequently, the atmospheric muon charge ratio $N_{\mu^{+}}/N_{\mu^{-}}$ is larger than unity.  A precise measurement of the atmospheric muon charge ratio can be used to constrain cosmic-ray shower models and, since atmospheric neutrinos are produced in conjunction with atmospheric muons, better determine atmospheric neutrino fluxes.\\

The differential muon production spectrum in extensive air showers can be parameterized as~\cite{Gaisser:1990vg}:
\begin{equation}
\frac{dN_{\mu}}{dE_{\mu}}\approx\frac{0.14\esurfmath^{-2.7}}{\textrm{cm}^{2}\textrm{ s }\textrm{sr GeV}}\times  \left( \frac{1.0}{1+\frac{\textrm{1.1}\ecossurfmath}{\epsilonmath_{\pi}}}+\frac{0.054}{1+\frac{\textrm{1.1}\ecossurfmath}{\epsilonmath_{K}}}\right ),
\label{eq:MuonFlux}
\end{equation}
where $E_{\mu}$ is the muon surface energy and $\theta$* is the zenith angle at the muon production point.  Accounting for the curvature of the earth, this angle is geometrically related to the observed zenith angle $\theta$,
\begin{equation}
\costhetamath=\sqrt{1-\textrm{sin}^{2}\theta\left ( \frac{R_{e}}{R_{e}+h}\right)^{2}},
\label{eq:StarAngle}
\end{equation}
where $R_{e}$ is the radius of the Earth, and $h$ is the muon production height for horizontal muons. Assuming the mean column depth for the initial cosmic ray primary interaction to be \unit[85]{g/cm$^{2}$} gives $h$=\unit[30]{km}\cite{Lipari:1993hd}.  The two terms in Eq.~(\ref{eq:MuonFlux}) represent the contribution to muon production from pion and kaon decay respectively.  The values $\epsilon_{\pi}$=\unit[115]{GeV} and $\epsilon_{K}$=\unit[850]{GeV} are the critical energies at the muon production height above which the pion and kaon interaction probability exceeds the decay probability.  The larger value of $\epsilon_{K}$ implies that the kaon contribution to the muon flux, and therefore to the charge ratio, will be more significant at values of the vertical muon surface energy \ecossurfspace exceeding $\epsilon_{\pi}$.  The charge ratio is expected to increase as a consequence of the increasing kaon contribution to the muon flux because in high energy interactions, single $K^{{}+{}}$'s can be produced in associated production with strange baryons while single $K^{{}-{}}$'s cannot.  The muon charge ratio is expected to decrease at even higher energies due to heavy flavor production~\cite{Zas:1992ci,Gaisser:2002jj} and changes in composition of the cosmic-ray primaries~\cite{Aglietta:2004ws,Derbina:2005ta,Schreiner:2009zz}.  However, as these latter two processes are expected to only affect the charge ratio at energies greater than those studied here they are not considered further. \\

Atmospheric muon charge ratio measurements utilizing deep underground detectors are typically higher in value than those produced using detectors which are shallower since they sample muons with a larger value of \ecossurf.  A previous measurement of the muon charge ratio utilizing the MINOS Far Detector, at a depth of \unit[2070]{} meters of water equivalent~(mwe), reported a value of $1.374 \pm 0.004$(stat.)$^{+0.012}_{-0.010}$(syst.)~\cite{Adamson:2007ww} for surface energies greater than 1 TeV.  Another measurement~\cite{:2010fg} of the charge ratio at TeV energies, conducted using the OPERA experiment located at a depth of \unit[3800]{mwe}, is in good agreement with the MINOS Far Detector result.  Atmospheric muon charge ratio measurements performed by the  L3+C~\cite{Achard:2004ws}, Bess-TEV~\cite{Haino:2004nq} and CMS~\cite{Khachatryan:2010mw} collaborations, for muons with surface energy \esurfspace above \unit[10]{GeV} and below \unit[300]{GeV}, are consistent with the 2001 world average of 1.268$\pm[0.008+0.0002\cdot\frac{\esurfmath}{\textrm{GeV}}]$~\cite{Hebbeker:2001dn}. \\

The MINOS Near and Far Detectors~\cite{Michael:2008bc} are large underground magnetic calorimeters at depths of \unit[225]{mwe} and \unit[2070]{mwe} respectively.  The detectors are designed to study neutrino interactions from the Fermilab Neutrinos at the Main Injector~(NuMI) beam~\cite{Crane:1995ky}, but also trigger on atmospheric muons between beam spills.  The depths of the MINOS detectors are ideal for making a measurement of the muon charge ratio at values of \ecossurfspace in the region dominated by the pion contribution and the transition region where the kaon contribution becomes significant. In this paper we present a measurement of the atmospheric muon charge ratio using the MINOS Near Detector. This result is then compared to the same measurement performed using the MINOS Far Detector~\cite{Adamson:2007ww}.\\

This paper is organized as follows. Sections ~\ref{sec:MINOSDETECTOR} and \ref{sec:MonteCarlo} describe the MINOS Near Detector and the atmospheric muon Monte Carlo simulation respectively.  Section~\ref{sec:EventSelection} describes the criteria for selecting atmospheric muon tracks with correct charge sign identification.  Section~\ref{sec:CombineFields} outlines the technique used to combine the data collected from two detector magnetic field configurations to determine the atmospheric muon charge ratio.  Section~\ref{sec:CombineFields} also elaborates on the evaluation of the systematic uncertainties.  Finally, Sec.~\ref{sec:TrueChargeRatio} presents the results of this analysis.\\

\section{The MINOS Near Detector}
\label{sec:MINOSDETECTOR}

The MINOS Near Detector~\cite{Michael:2008bc} at Fermilab is a magnetized-steel and scintillator sampling calorimeter.  It is located \unit[94]{m} underground, with an approximately flat overburden of \unit[225]{mwe}.  Its dimensions are \unit[3.8]{m}$\times$\unit[4.8]{m}$\times$\unit[16.6]{m}.  The detector contains 282 vertical steel planes, each \unit[2.54]{cm} thick.  The scintillator layers are composed of either 64 or 96, \unit[4.1]{cm} wide and \unit[1]{cm} thick strips which vary in length from 2.5 to \unit[4]{m}.  The scintillator strips are rotated by 90\ensuremath{^\circ} with respect to the previous layer to allow for three dimensional track reconstruction.  The first 120 steel planes each have a scintillator layer attached to their surface.  The scintillator layer of every fifth steel plane contains 96 strips, and is said to be fully instrumented.  The remaining four planes in each set of five are partially instrumented and contain 64 strips.  Of the last 162 steel planes, only every fifth plane has an attached scintillator layer.  These scintillator layers are fully instrumented. \\

Scintillation light is collected by  wavelength shifting fibers embedded in the scintillator strips.  Each strip is coupled to a single pixel on a 64-pixel multi-anode photo-multiplier tube~\cite{Tagg:2004bu} by a clear fiber.  Each PMT pixel is digitized continuously at a frequency of \unit[53]{MHz}. The detector response for a candidate atmospheric muon event is recorded when either four strips in five sequential planes, or when strips from any 20 planes, register a signal above the 1/3 photo-electron dynode threshold within \unit[151]{ns}.\\
    
 The detector's approximately toroidal magnetic field~\cite{Michael:2008bc} varies in strength from \unit[2.1]{T} near the magnetic coil hole to \unit[0.2]{T} near the periphery of the steel planes. The magnetic field can be oriented to focus either northerly going $\mu^{{}-{}}$ or $\mu^{{}+{}}$. These magnetic field orientations will be referred to as ``forward'' and ``reverse'', respectively.  The detector is oriented horizontally and its long axis points 26.5548$^{\circ}$ west of north.  The curvature induced by the magnetic field together with the three dimensional track reconstruction allows the determination of the charge sign of a muon traversing the detector. \\
 
Detailed information regarding the MINOS Near Detector, the electronics and the data acquisition systems can be found in~\cite{Michael:2008bc,Belias:2004bj,Cundiff:2006yz}. \\

\section{Simulated Atmospheric Muons}
\label{sec:MonteCarlo}
A sample of 42$\times$10$^{6}$ simulated atmospheric muon events are used to evaluate the analysis sensitivities and efficiencies, as well as to help assess systematic errors. Surface level muon events were generated by the HEMAS~\cite{Scapparone:1998rx} atmospheric cascade simulation.  Muons expected to intersect with the detector are propagated through a GEANT3~\cite{Agostinelli:2002hh} model of the overburden and the MINOS Near Detector. \\

\section{Event Selection}
\label{sec:EventSelection}
\label{sec:preanalysiscuts}

The selection criteria have been chosen to optimize the event selection efficiency observed in the data and the charge sign identification purity obtained from the Monte Carlo simulations.  The data sample for this analysis consists of atmospheric muons with well-reconstructed energy and charge sign.  Pre-selections are made to reject events that are not consistent with the passage of an atmospheric muon through the detector.  Further selections are applied to ensure track reconstruction quality and good charge sign determination.  A summary of the selection criteria, as well as the selection efficiencies and charge identification purities, can be found in Table~\ref{tab:FullQualityTable}. \\

This analysis uses data collected between 2006 and 2009. During the data taking period, the direction of the magnetic field was periodically changed between the two orientations referred to as forward and reverse magnetic field. The final reverse magnetic field exposure was \unit[150.5]{days}. Equivalent exposures of forward magnetic field data were collected immediately following the reverse magnetic field data periods to reduce systematic uncertainties.  A total of 7.16$\times$10$^{8}$ triggers were collected over a combined \unit[301]{day} exposure. \\

\subsection{Pre-Selections and Track Quality }
\label{sec:trackquality}

For an event to be included in this analysis it must possess a single downward-going atmospheric muon track, have been collected during a period of good detector run conditions, and have a reconstructed initial interaction point~(RIIP) within \unit[50]{cm} of the detector edge.  If the reconstructed initial interaction point is further than \unit[3]{cm} outside the detector volume the track is rejected.  The curvature of the track is known to be poorly determined, and the track is rejected, if any scintillator strips hit are further than \unit[3]{cm} from the reconstructed position of the track, or if the track does not pass through a region of the detector where there is scintillator on each layer of steel. \\

 Track reconstruction errors may occur when the event contains a large amount of activity which is not related to the track.  These extra hits degrade the charge sign determination and can be generated by muon bremsstrahlung, natural radioactivity, or by electrical and optical cross-talk between the channels on the multi-anode PMT~\cite{Tagg:2004bu}.  Events are rejected if more than 40\% of the strips hit are not track-related. Muon tracks determined to be poorly reconstructed by internal consistency checks of the reconstruction algorithm are also removed from the data sample.  \\

\subsection{Charge Sign Quality Selection}
\label{sec:qconfidencecuts}  

\begin{figure}[h!tb]
\begin{center}
\includegraphics[width=0.5\textwidth]{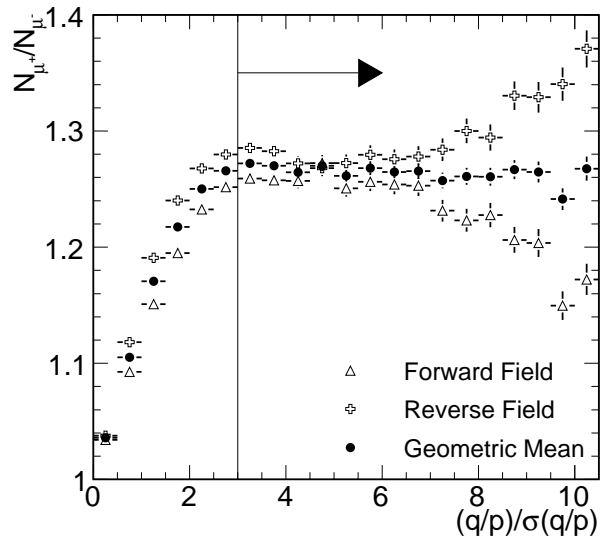}
\caption{Charge ratio as a function of $(q/p)/\sigma(q/p)$ after all selections and requiring that $BdL>\unit[3.0]{~T\cdot m}$. The vertical line is the $(q/p)/\sigma(q/p)$ threshold value used in the charge sign quality selection. } 
\label{fig:CRvsQPplot}
\end{center}
\end{figure}

\begin{figure}[h!tb]
\begin{center}
\includegraphics[width=0.5\textwidth]{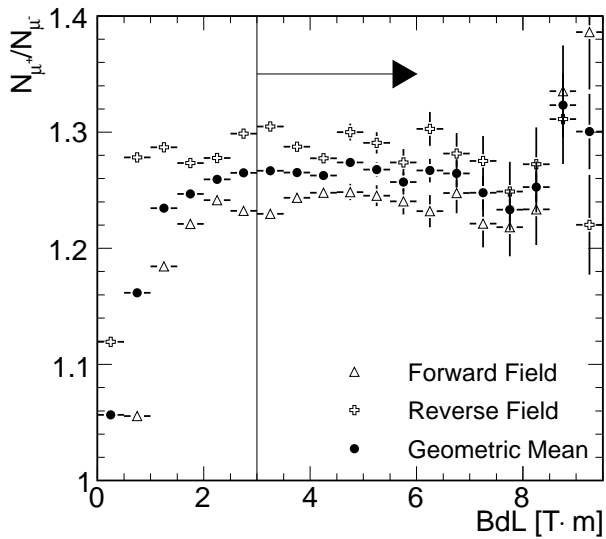}
\caption{Charge ratio as a function of {\it BdL}, after all selections and requiring that $(q/p)/\sigma(q/p)>3.0$.  The vertical line is the {\it BdL} threshold value used in the charge sign quality selection.}
\label{fig:CRvsBDLplot}
\end{center}
\end{figure}

 Two selection variables are used to further increase the degree of confidence in the assigned curvature and charge sign of the track.  The Kalman filter~\cite{Fruhwirth:1987fm} used in the track curvature fitting provides an uncertainty $\sigma{(q/p)}$ on the measured value of $q/p$, where $q$ is the charge and $p$ the momentum of the track.  The first charge sign quality selection is based on the value of $(q/p)/\sigma{(q/p)}$ determined by the track fitter.  The second selection variable {\it BdL} is defined to be equivalent to $\sum\limits_{i=1}^{N}|B_{i}\times dL_{i}|$ where {\it N} is the total number of planes in the muon track, and {\it B$_{i}$} and {\it dL$_{i}$} are the magnetic field and muon path length vectors, respectively, at plane {\it i}.  A selection based on {\it BdL} is used to ensure that the magnitude of the bending due to curvature is larger than the apparent bending due to multiple scattering.  Figure~\ref{fig:CRvsQPplot} shows the muon charge ratio as a function of $(q/p)/\sigma(q/p)$ and Fig.~\ref{fig:CRvsBDLplot} the charge ratio as a function of $BdL$, for data collected in both magnetic field orientations as well as the combined data set. The charge ratio for data collected during a single magnetic field orientation is defined as the ratio of positive to negative muons collected only in that orientation.  The observed variation in the charge ratio above the selection thresholds in a single field orientation, as well as the difference in the charge ratio between the two different field orientations, stems from acceptance effects due to the magnetic field, detector asymmetry and detector alignment errors.  The technique used to combine data taken in the two field orientations and remove these biases is discussed in the next section. For this analysis we required that $BdL>\unit[3.0]{~T\cdot m}$ and $(q/p)/\sigma(q/p)>3.0$.  Below these values the charge ratio tends towards unity, indicating a degradation of the charge sign determination.  Events which have passed all the selections described in this section are used in the calculation of the atmospheric muon charge ratio described in the next section.\\ 

\begin{table*}[t!h]
\begin{tabular}{|l|c|c|} \hline \hline
& Data           & Monte-Carlo \\
& Selection      & Charge ID \\
Selection          & Efficiency     & Purity \\ \hline \hline
Number of Triggers              & 7.16$\times 10^{8}$ (100\%)           & - \\ \hline \hline
Pre-Selections                 & \multicolumn{2}{|c|}{} \\  \hline
Single Track Events            & 3.18$\times 10^{8}$ (44.46\%)         & 70.7\% \\
Detector Quality               & 3.16$\times 10^{8}$ (44.19\%)         & 70.7\% \\ \hline\hline
%Entrance Vertex                & 2.39$\times 10^{8}$ (33.44\%)         & 70.5\% \\ \hline \hline
Track Quality Selections       & \multicolumn{2}{|c|}{}   \\ \hline
RIIP and Curvature Selections    & 1.01$\times 10^{8}$ (14.19\%) & 72.9\%  \\ 
Track-related Activity   & 6.13$\times 10^{7}$ (8.56\%)         & 77.5\% \\
Good Reconstruction            & 5.31$\times 10^{7}$ (7.42\%)         & 78.8\% \\ \hline \hline
Charge Sign Quality Selections   & \multicolumn{2}{|c|}{}   \\ \hline
$(q/p)/\sigma(q/p)>3.0$        & 1.11$\times 10^{7}$ (1.55\%)         & 97.0\% \\
{\it BdL} $>$3.0 T$\cdot$m     & 3.23$\times 10^{6}$ (0.45\%)         & 99.5\% \\ \hline \hline
\end{tabular}
\caption{Summary of the event selection.  Each row shows the total number of events remaining after all the applied cuts; in parenthesis, the percentage of events remaining; and lastly, the percentage of the remaining Monte Carlo events which have their charge sign determined correctly.}
\label{tab:FullQualityTable}
\end{table*}

\section{Charge Ratio Determination}
\label{sec:CombineFields}
\label{sec:geomean}
\begin{figure*}[htb]
\begin{center}
\includegraphics[width=1.0\textwidth]{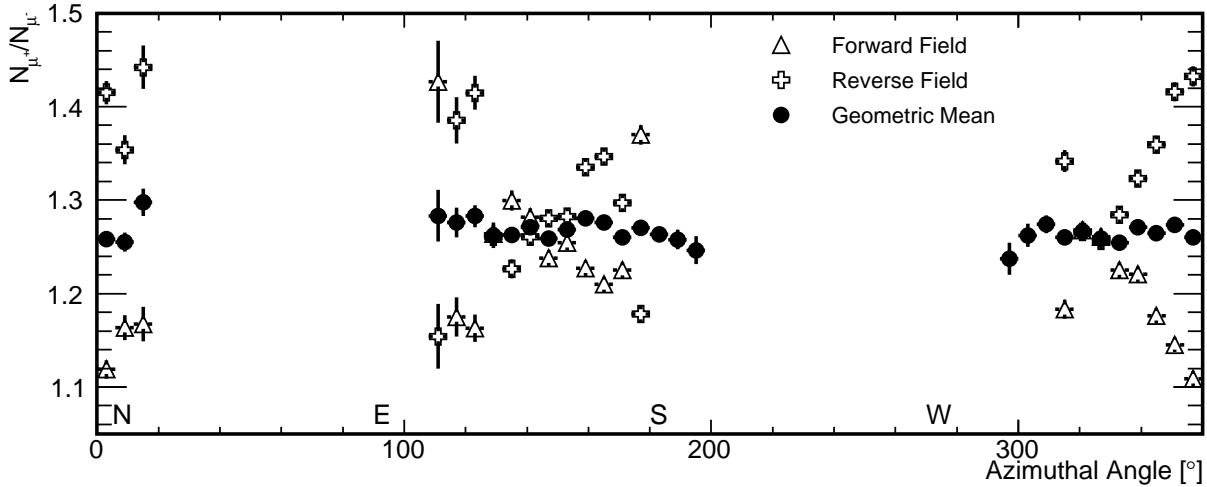}  
\caption{The observed charge ratio as a function of azimuthal angle.  The charge ratio varies as a function of azimuthal angle in the forward and reversed data sets due to acceptance effects and alignment errors.  When data from the two fields are combined using Eq.~(\ref{eq:finaleq}) a flat distribution is obtained, indicating that the residual uncertainty due to these biases is small.  The uncertainties shown are statistical only. }
\label{fig:CRvsAzimuth}
\end{center}
\end{figure*}

As demonstrated in the previous section acceptance effects in the Near Detector introduce a  bias in the charge ratio when it is calculated using only data from a single magnetic field orientation.  Figure~\ref{fig:CRvsAzimuth} shows the charge ratio as a function of azimuthal angle, a variable sensitive to these biases, for the forward and reverse magnetic field data sets.  Canceling these biases is done in the same manner as described in ~\cite{Adamson:2007ww,Matsuno:1984kq}. If $\varepsilon_{1}$ is the efficiency for the selection of $\mu^{{}+{}}$ and $\varepsilon_{2}$ is the selection efficiency of $\mu^{{}-{}}$ in the forward field direction (FF) then the selection efficiencies for  $\mu^{{}+{}}$ and  $\mu^{{}-{}}$ in the reverse field direction (RF) are  $\varepsilon_{2}$ and $\varepsilon_{1}$ respectively. Two independent equations for the charge ratio, $r_{a}$ and $r_{b}$, can be constructed in which the acceptance effects cancel. These ratios, corrected for live time, are 
\begin{equation}
r_{a}=(N^{\mu^{{}+{}}}_{FF}/t_{FF})/(N^{\mu^{{}-{}}}_{RF}/t_{RF}),
\label{eq:ra}
\end{equation}
and
\begin{equation}
r_{b}=(N^{\mu^{{}+{}}}_{RF}/t_{RF})/(N^{\mu^{{}-{}}}_{FF}/t_{FF}),
\label{eq:rb}
\end{equation}
where $N^{\mu^{{}+{}}}$ ($N^{\mu^{{}-{}}}$) is the number of positive (negative) muons observed during an exposure time {\it t} in a particular magnetic field orientation.  The geometric mean of $r_{a}$ and $r_{b}$
\begin{equation}
\frac{N_{\mu^{{}+{}}}}{N_{\mu^{{}-{}}}}=\sqrt{r_{a}r_{b}}=\sqrt{\left (\frac{N^{\mu^{{}+{}}}_{FF}}{N^{\mu^{{}
-{}}}_{FF}} \right ) \left (\frac{N^{\mu^{{}+{}}}_{RF}}{N^{\mu^{{}-{}}}_{RF}} \right ) },
\label{eq:finaleq}
\end{equation}
provides a measurement of the charge ratio that is free of biases due to geometric acceptance effects, alignment errors and the different magnetic field live times.\\

Figure~\ref{fig:CRvsAzimuth} illustrates that the significant bias in the charge ratio, that is apparent in a single field orientation data set, is strongly suppressed in the geometric mean.\\
 
After the selection criteria in Table~\ref{tab:FullQualityTable} have been applied we obtain a final data set of 3,234,066 events.  In the forward field sample we select 893,854 $\mu^{+}$ and 721,428 $\mu^{-}$.  In the reverse field sample we select 912,944 $\mu^{+}$ and 705,840 $\mu^{-}$.  The resulting charge ratio obtained by applying Eq.~(\ref{eq:finaleq}) is $N_{\mu^{{}+{}}}/N_{\mu^{{}-{}}}$=1.266$\pm$0.001(stat.).\\

\subsection{Systematic Uncertainties}
\label{sec:ErrorDiscussion}
The event selection criteria are chosen to remove events in which the charge sign of the muon track has been incorrectly assigned.  Systematic uncertainties associated with event selection are determined by establishing the sensitivity of the measured charge ratio to variations in the selections above their thresholds. Two selections, track-related activity and $(q/p)/\sigma(q/p)$, display variations in the observed charge ratio above their thresholds. These data were divided into three equal statistics samples, corresponding to increasing confidence in the muon charge sign, and their charge ratios calculated.  The maximum deviation from the nominal charge ratio, 0.012 and 0.006 for the track related activity and $(q/p)/\sigma(q/p)$ selections respectively, is taken as the systematic uncertainty associated with that particular selection.\\

\begin{figure}[htb]
\begin{center}
\includegraphics[width=0.5\textwidth]{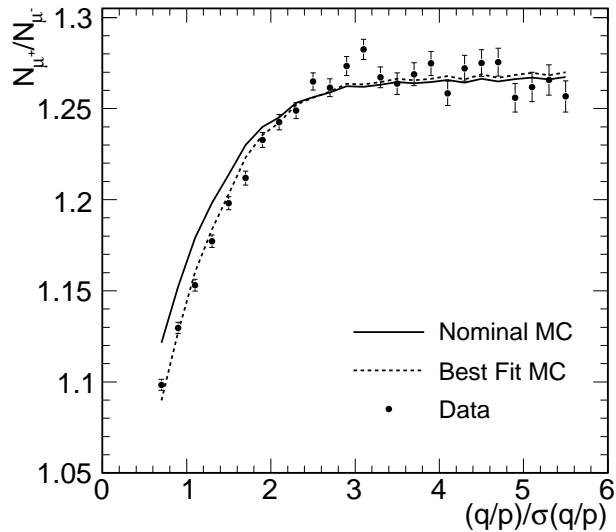}  
\caption{Charge ratio as a function of $(q/p)/\sigma(q/p)$, compared with the nominal and best-fit charge misidentification Monte Carlo simulations.}
\label{fig:MCtweak}
\end{center}
\end{figure}

Another systematic uncertainty relates to the remaining events that have a misidentified charge sign after the selections. Since the atmospheric muon charge ratio is a value greater than unity more positive than negative muons will have their charge sign misidentified. Thus charge sign misidentification can only decrease the measured charge ratio resulting in a positive one-sided systematic uncertainty.  Monte Carlo studies suggest that 0.5\% of the events in the final data  sample have an incorrect charge sign determination.  However, one cannot discount the possibility that the misidentification rate is different in the data than in the Monte Carlo simulation.  For this analysis it is assumed that the true charge sign misidentification rate in data differs from that in the nominal Monte Carlo by an energy independent factor $\alpha$.  The true charge sign misidentification rate in data can be estimated by exploiting the fact that it influences the shape of the charge ratio versus $(q/p)/\sigma(q/p)$ curve shown in Fig.~\ref{fig:CRvsQPplot}. A toy Monte Carlo was written to produce similar curves assuming a charge ratio R$_{a}$, and a charge sign misidentification rate equivalent to that in the nominal Monte Carlo scaled by a factor $\alpha$. The difference between the measured charge ratio and R$_{a}$ for the values of $\alpha$ and R$_{a}$ that provide the best agreement between the data and the Monte Carlo prediction is taken as a systematic uncertainty on the measured charge ratio.  Figure~\ref{fig:MCtweak} plots the charge ratio as a function of $(q/p)/\sigma(q/p)$, along with the nominal and best-fit Monte Carlo.  The best fit Monte Carlo is obtained using a charge ratio $R_{a}$ of 1.272. A similar study was performed on the curvature of the charge ratio versus $BdL$ data in Fig.~\ref{fig:CRvsBDLplot}; in that study the best fit Monte Carlo is obtained with an actual charge ratio of 1.266. The maximum deviation of R$_{a}$ from the nominal charge ratio is +0.006, which is taken as the one-sided systematic uncertainty associated with charge sign misidentification uncertainties.\\   

Two alternative charge confidence selection criteria, similar in motivation to the $BdL$, have been investigated. The first alternative required that the muon traverse at least 27 planes at positions covered by the partially instrumented scintillator planes, within \unit[1.75]{m} of the magnetic coil hole.  The second alternative required that at least 37 planes be traversed where the $|B\times dL|$ for each of those planes is greater than \unit[0.03]{T$\cdot$m}. These particular selection criteria have been chosen to optimize data selection efficiency while maintaining a charge misidentification rate in the Monte Carlo similar to that of the default $BdL$ selection.  The maximum deviation from the nominal charge ratio was observed to be 0.003. This value is taken as the systematic uncertainty associated with the $BdL$ selection. \\

An imperfect reversal of the detector's magnetic field would introduce a geometric bias and lead to a  systematic uncertainty in the charge ratio stemming from imperfect acceptance asymmetry cancellation. This error can be determined by evaluating the charge ratios calculated using Eq.~(\ref{eq:ra}) and Eq.~(\ref{eq:rb}) since the quantities $r_{a}$ and $r_{b}$ will diverge as the magnitude of this bias increases.  It is found that $r_{a}$ and $r_{b}$ agree to within their statistical uncertainties, indicating that the magnetic field bias is negligible compared to the statistical uncertainty.\\
%%%%%%%%%%%%%%%%%%%%%%%%%%%

\begin{table}
\begin{tabular}{|l|c|} \hline
Uncertainty Classification     & $\Delta\frac{N_{\mu^{{}+{}}}}{N_{\mu^{{}-{}}}}$ \\ \hline
Track-Related Activity   & $\pm$0.012 \\
$(q/p)/\sigma(q/p)$      & $\pm$0.006 \\
$BdL$ Selection       & $\pm$0.003 \\
Charge Misidentification & +0.006 \\ \hline
 Total Systematic Uncertainty       & $^{+0.015}_{-0.014}$   \\ \hline
\end{tabular}
\caption{Summary of the systematic uncertainties in this analysis. The total systematic uncertainty is the quadratic sum of the individual uncertainties  The magnitude of these uncertainties are independent of \esurfspace and \ecossurf.}
\label{tab:SystematicSummary}
\end{table}

Table~\ref{tab:SystematicSummary} lists the systematic errors that are discussed above, and their assigned values.  Including systematic uncertainties the atmospheric muon charge ratio measured at the MINOS Near Detector is 1.266$\pm$0.001(stat.)$^{+0.015}_{-0.014}$(syst.). \\

\subsection{The Muon Charge Ratio Underground}

\begin{figure}[htb]
\begin{center}
\includegraphics[width=0.5\textwidth]{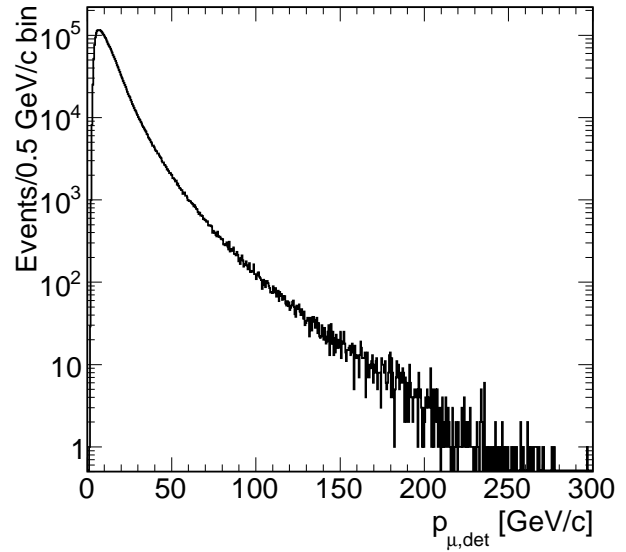}
\caption{The reconstructed momentum of the muon tracks which survive all the selection criteria outlined in Table~\ref{tab:FullQualityTable}.  The mean momentum is \unit[15]{GeV}. }
\label{fig:1DmomentumDistribution}
\end{center}
\end{figure}

\begin{figure}[htb]
\begin{center}
\includegraphics[width=0.5\textwidth]{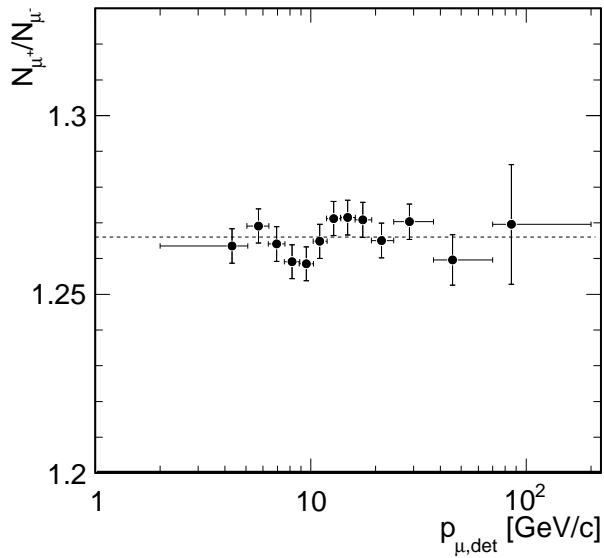}
\caption{Charge ratio as a function of reconstructed underground track momentum. The y-axis uncertainties are statistical. The x-axis uncertainties are the width of the momentum bins used; the data are plotted at the median momentum values.  The dotted horizontal line is the best fit charge ratio of 1.266.}
\label{fig:1Dmomentum}
\end{center}
\end{figure}

 The maximum momentum for which the charge sign of a muon track can accurately be determined with the MINOS Near Detector is limited by the scintillator granularity and the strength of the magnetic field.  The reconstructed momenta $p_{\mu,det}$ of all muon tracks which pass the selections outlined in Table~\ref{tab:FullQualityTable} is shown in Fig.~\ref{fig:1DmomentumDistribution}.   The maximum momentum of muons used in this analysis is approximately \unit[300]{GeV}; the mean momentum is  \unit[15]{GeV}.  The observed charge ratio as a function of reconstructed track momentum underground is shown in Fig.~\ref{fig:1Dmomentum}.  The charge ratio is observed to be flat as a function of track momentum to within the statistical uncertainty of the measurements. \\

\section{The Atmospheric Muon Charge Ratio at the Surface}
\label{sec:TrueChargeRatio}

\label{sec:InducedMuons}
\label{sec:ToSurface}
The muon energies measured at the Near Detector depth must now be converted to energies at the surface of the Earth by accounting for energy lost by the muons in the overburden above the detector. For muons, the energy loss in matter can be parameterized by
\begin{equation}
-\frac{dE}{dX}=a(E_{\mu})+\sum_{n=1}^{3}b_{n}(E_{\mu})E_{\mu},
\label{eq:EnergyLoss}
\end{equation}
where X is the slant depth; {\it a} is the ionization term and the {\it b$_{n}$} account for the radiative energy loss from bremsstrahlung, pair production and photo-nuclear processes.  These parameters have a mild energy dependence, whose values for  standard rock can be found in~\cite{Yao:2006px}.  Using the technique outlined in~\cite{Reichenbacher:2007dm}, the energy lost by muons traversing the Near Detector overburden has been calculated as a function of reconstructed track energy and zenith angle.  These energies have been calculated assuming a flat vertical overburden of \unit[224.6]{mwe}, which is comprised of two distinct geological layers.  Above the cavern hall lies \unit[72.1]{m} of Dolomite/Shale bedrock with a density of \unit[2.41]{g/cm$^{3}$}, followed by \unit[22.2]{m} of Glacial till with a density of \unit[2.29]{g/cm$^{3}$}.  It has been suggested in \cite{Schreiner:2009zz} that at the MINOS Near Detector depth the slightly higher rate of energy loss of $\mu^{+}$ over $\mu^{-}$ \cite{Smith:1953zz,Barkas:1956zz,Jackson:1972zz,Clark:1972vh,Jackson:1998jq,Lee:2004vb} could lead to a surface charge ratio that is slightly higher than that observed underground.  However, as the magnitude of this effect is negligible when compared to the systematic uncertainties of this measurement we have assumed the same energy loss function for both charges. \\

 The surface energy resolution obtained with this extrapolation method is dependent on the accuracy of the overburden model and the muon pointing accuracy.  Topographical measurements \cite{TopoMaps} show that the surface altitude varies by no more than \unit[13]{m} within \unit[3]{km} of the detector.  This contributes a 14\% uncertainty on the surface energy estimation at all zenith angles.  The zenith pointing resolution of the Near Detector has been determined to be 1.1$\pm$0.2$^{\circ}$ by measuring the zenith angle separation between the two muons in multi-muon cosmic-triggered events for which both muons pass the charge ratio selection criteria. The corresponding error on the surface energy is negligible for vertical muons increasing to 5\% at 70$^{\circ}$,14\% at 81$^{\circ}$, 25\% at 85$^{\circ}$ and 50\% at 87$^{\circ}$. \\

\begin{figure}[h!tb]
\begin{center}
\includegraphics[width=0.5\textwidth]{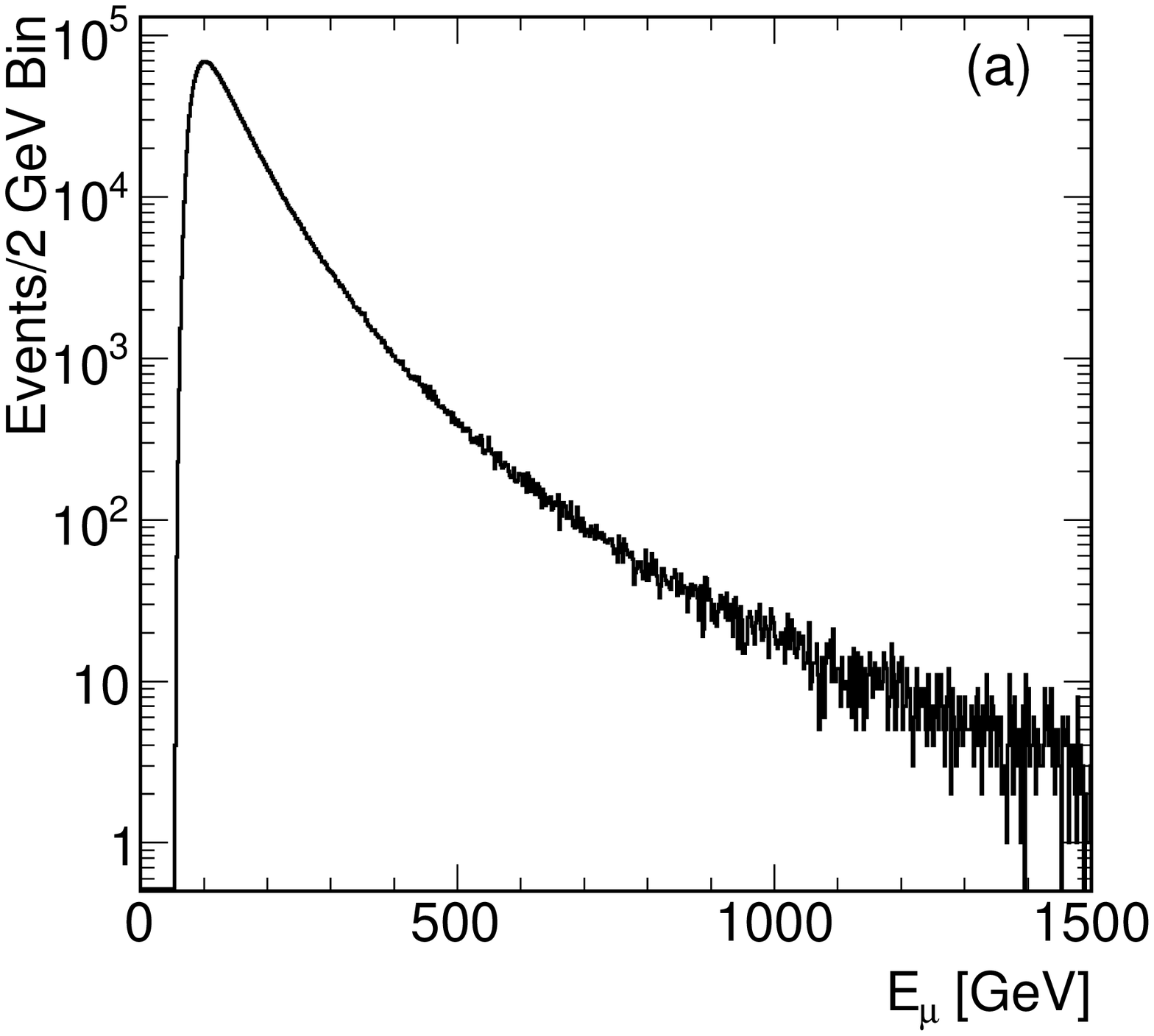}
\includegraphics[width=0.5\textwidth]{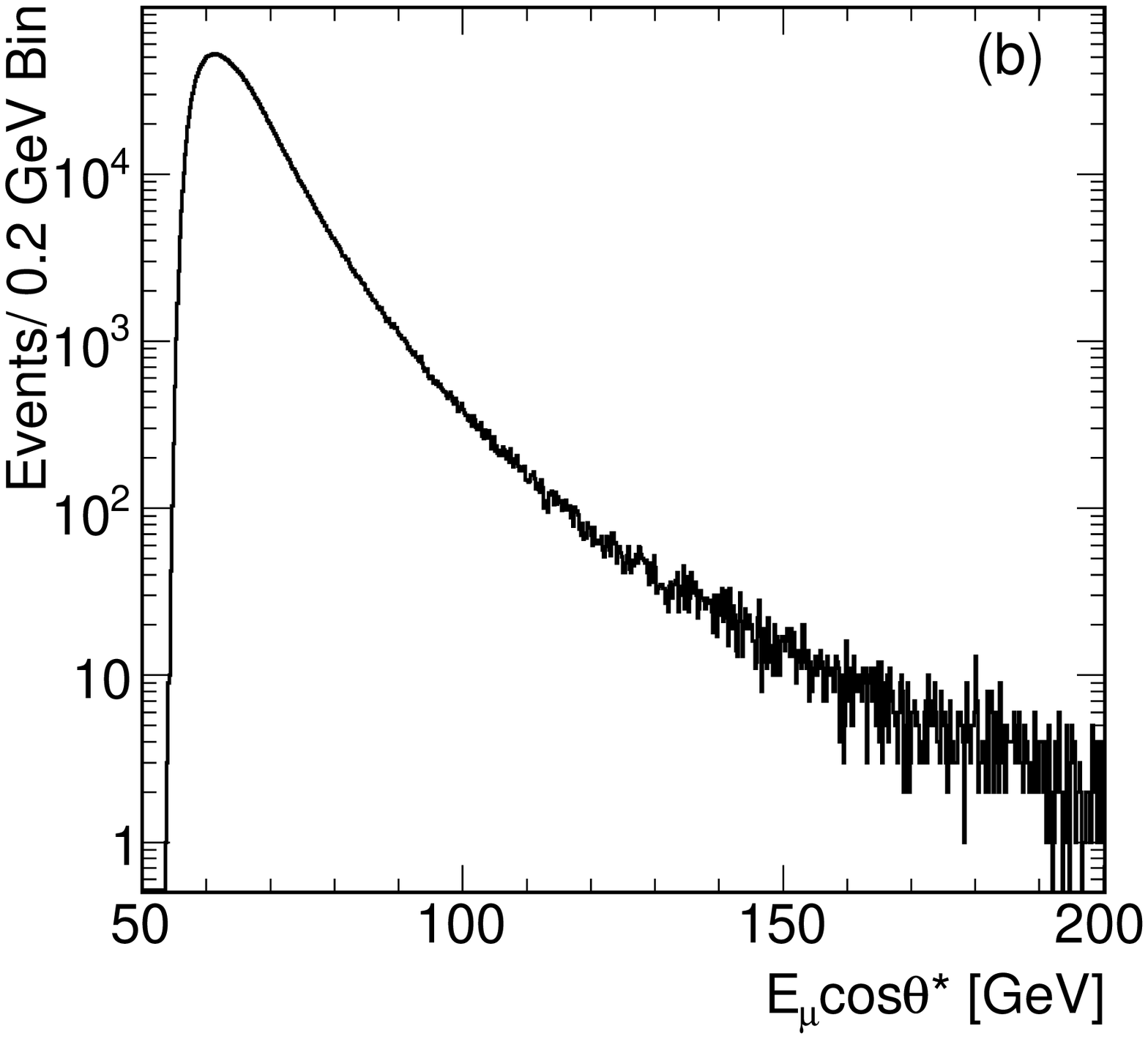}
\caption{The extrapolated (a) muon surface energy and (b) \ecossurfspace distributions of muons passing all the analysis selection criteria.  The mean surface energy is \unit[152]{GeV} and the mean \ecossurfspace is \unit[66]{GeV}. }
\label{fig:EnergyDistribution}
\end{center}
\end{figure}

 The extrapolated muon surface energy distribution, \esurf, is plotted in Fig.~\ref{fig:EnergyDistribution}(a).  The muons which populate the high energy tail have a large zenith angle and thus pass through the largest amount of matter before reaching the detector.  To first approximation the minimum energy a surface muon needs to reach the Near Detector is \unit[52]{GeV}/\oldcostheta.  The \ecossurfspace distribution is much narrower than the \esurfspace distribution as the \costhetaspace and \oldcosthetaspace terms effectively cancel.  Due to this cancellation the muon events which occupy the high energy tail in Fig.~\ref{fig:EnergyDistribution}(b) are those which possess the highest reconstructed momentum at the detector. \\

% \afterpage{\clearpage}

\begin{table*}
\begin{tabular}{|c|c||c|c||c|c||c|c||} \hline
\multicolumn{2}{|c||}{\costheta}  & \multicolumn{2}{c||}{Surface Energy} & \multicolumn{2}{c||}{\ecossurf}  & \multicolumn{2}{c||}{Charge Ratio }\\ 
\multicolumn{2}{|c||}{}  & \multicolumn{2}{c||}{[GeV]}  & \multicolumn{2}{c||}{[GeV]} &  \multicolumn{2}{c||}{$^{+0.015}_{-0.014}(syst.)$}             \\ 
Interval  & Median   & Interval       & Median   & Interval & Median & r & $\pm$(stat.) \\ \hline
$>$0.9   &	0.914&	55-67   &	61.5&	53.5-61.0&	56.5&	1.257&	0.069\\
0.8-0.9   &	0.824&	60-77   &	70.5&	54.2-62.9&	58.0&	1.270&	0.013\\
0.7-0.8   &	0.735&	69-91   &	80.5&	54.6-65.0&	59.2&	1.277&	0.005\\
0.6-0.7   &	0.642&	79-112  &	94.5&	55.2-69.0&	60.8&	1.270&	0.003\\
0.5-0.6   &	0.549&	95-142  &	115 &	56.2-74.0&	62.7&	1.269&	0.003\\
0.4-0.5   &	0.452&	115-191 &	145 &	57.4-80.9&	64.8&	1.259&	0.003\\
0.3-0.4   &	0.356&	148-280 &	195 &	59.4-91.0&	68.0&	1.263&	0.003\\
0.2-0.3   &	0.262&	212-484 &	290 &	63-109   &	75  &	1.263&	0.005\\
0.15-0.2  &	0.181&	382-864 &	513 &	76-140   &	92  &	1.276&	0.013\\
$<$0.15 &	0.139&	697-3450&	964 &	104-390  &	134 &	1.249&	0.032\\ \hline
\end{tabular}
\caption{The charge ratio, \esurfspace and \ecossurf, in equal bins of \costheta. The range of energies observed in each \costhetaspace bin is comparable to the energy resolution expected from the surface extrapolation.  The charge ratio is observed to be independent of \costheta, \esurfspace and \ecossurfspace to within the uncertainties of our measurements. }  
\label{tab:CR_vs_cos_vs_E}
\end{table*}

Table~\ref{tab:CR_vs_cos_vs_E} presents the measured atmospheric muon charge ratio, \esurfspace and \ecossurfspace in equal bins of \costheta.  The median surface energy increases as \costhetaspace decreases due to the increasingly large overburden. \ecossurfspace increases more slowly than \esurfspace as the increase is due primarily to the detector's larger analyzable momentum range as a function of zenith angle\cite{Goodman:2007}.   The charge ratio is observed to be independent of \costheta, \esurfspace and \ecossurfspace to within the uncertainties of our measurements.  The charge ratio measurement at \esurf$=$\unit[964]{GeV} is consistent with the MUTRON spectrograph measurement~\cite{Matsuno:1984kq}, but is lower than other TeV energy scale measurements~\cite{Ashley:1975uj,Adamson:2007ww,:2010fg}.  This difference is due to the range of zenith angles sampled by each of the detectors and the \costhetaspace dependency of Eq.~(\ref{eq:MuonFlux}).\\

The ``$\pi$K''~\cite{Adamson:2007ww,Schreiner:2009zz} model is derived from the differential muon production spectrum parameterization given in Eq.~(\ref{eq:MuonFlux}), and predicts that the muon charge ratio is only dependent on \ecossurf.  The $\pi$K model is a qualitative model describing the relative contribution of pions and kaons to the atmospheric muon charge ratio.  Following the prescription in~\cite{Adamson:2007ww} and defining $f_{\pi}$ and $f_{K}$ as the fraction of all decaying pions and kaons which decay with a detected $\mu^{{}+{}}$, the atmospheric muon charge ratio $N_{\mu^{{}+{}}}/N_{\mu^{{}-{}}}$ is given in Eq.~(\ref{eq:CRecostheta}):

\begin{equation}
\label{eq:CRecostheta}	
\rpm = \frac{\left \{ 
{\frac{f_\pi}{1 ~+~ {1.1 \ecc}/{\epsilon_\pi} } 
~+~\frac{0.054\times f_K} {1~+~{1.1 \ecc}/{\epsilon_K} }}
\right \} }
{\left \{\frac{1-f_\pi}{1~+~{1.1 \ecc}/{\epsilon_\pi} } 
~+~\frac{0.054\times(1- f_K)}  
{1~+~{1.1 \ecc}/{\epsilon_K}}\right \}}
\end{equation}
\\

\begin{figure*}[htb]
\begin{center}
\includegraphics[width=\textwidth]{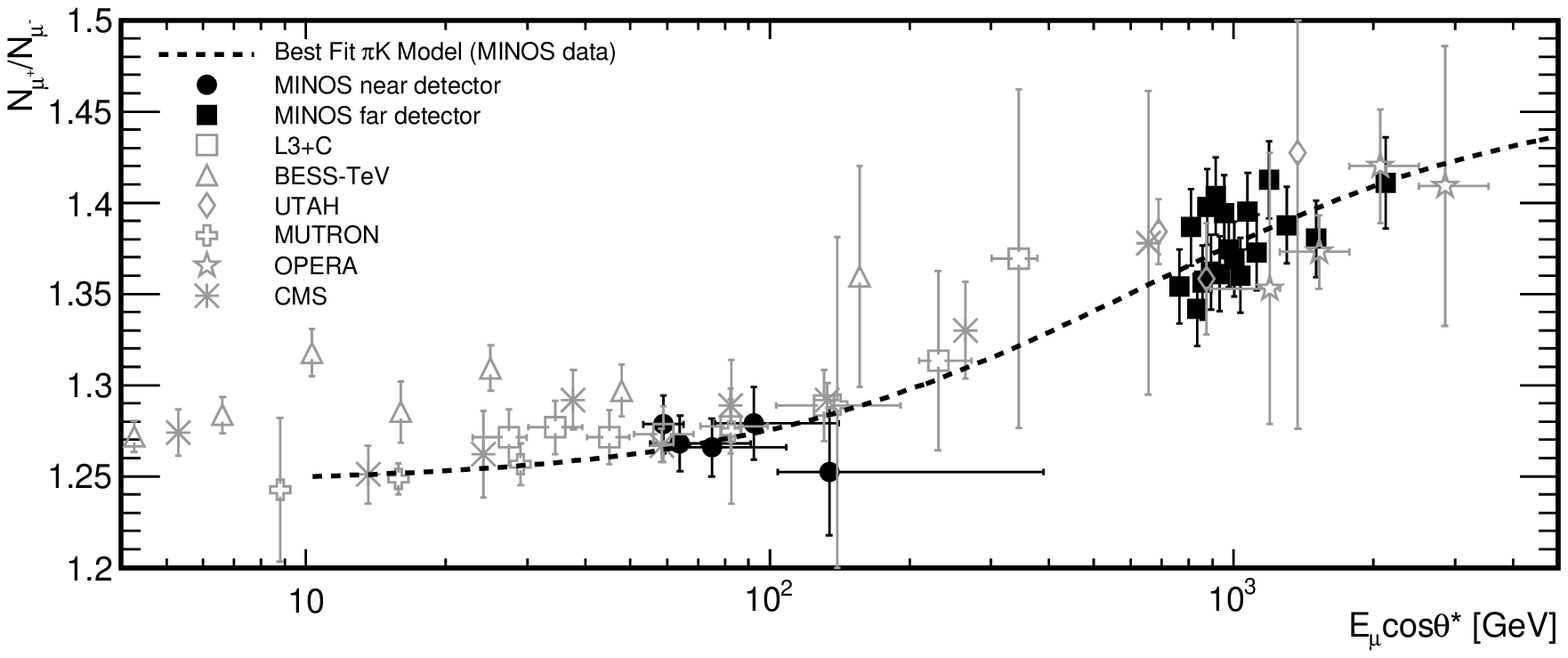}
\caption{The atmospheric muon charge ratio as a function of \ecossurf. The y-axis uncertainties are the statistical and systematic uncertainties added in quadrature.  The MINOS Near Detector data, re-binned in five equal \costhetaspace intervals, are plotted for each bin at the median \ecossurfspace values; the x-axis uncertainties are the bin widths. The dashed line is the best fit curve to the $\pi$K model using only MINOS Near and Far Detector data.}
\label{fig:CRecos}
\end{center}
\end{figure*}

The charge ratio as a function of \ecossurfspace is plotted in Fig.~\ref{fig:CRecos} along with the results from L3+C~\cite{Achard:2004ws}\cite{MauryNote}, Bess-TEV~\cite{Haino:2004nq}, UTAH~\cite{Ashley:1975uj}, MUTRON~\cite{Matsuno:1984kq}, OPERA~\cite{:2010fg}, CMS~\cite{Khachatryan:2010mw} and the MINOS Far Detector~\cite{Adamson:2007ww,Schreiner:2009zz}. In instances where the data were not already listed as a function of \ecossurfspace we convolved the published \oldcosthetaspace acceptances with the given \esurfspace data, with assistance from the original authors in the case of L3+C~\cite{L3Cemail}. \\

A $\chi^{2}$ per degree of freedom fit test to the $\pi$K model was performed over ($f_{K}$,$f_{\pi}$) space using only the charge ratio measurements from the MINOS Near and Far Detectors.  The $\chi^{2}$ minimum was found at $f_{\pi}$=0.55 and $f_{K}$=0.70.  These results are consistent with earlier fits by~\cite{Adamson:2007ww,Schreiner:2009zz,:2010fg}.  The best fit curve to the data is plotted in Fig.~\ref{fig:CRecos} and indicates that the kaon contribution to the charge ratio becomes significant at \ecossurfspace greater than a few hundred GeV.  Using two functionally identical detectors, at two different depths, we have demonstrated that the increase in charge ratio observed at the deeper Far Detector is consistent with an increase in the fraction of observed muons arising from kaon decays in the extensive air shower. 

\section{Summary}

A charge ratio measurement has been performed on 301 days of atmospheric muon data collected using the MINOS Near Detector.  The atmospheric muon charge ratio measured at \unit[225]{mwe} underground is
\begin{equation}
\frac{N_{\mu^{{}+{}}}}{N_{\mu^{{}-{}}}}=1.266 \pm 0.001(stat.)^{+0.015}_{-0.014} (syst.).
\end{equation}
The reconstructed underground energy and zenith angle for each muon used in the analysis were used to extrapolate the muon surface energy.  No statistically significant change was observed in the charge ratio as a function of either the surface energy \esurf, from \unit[62]{GeV} to \unit[960]{GeV}, or in the vertical muon surface energy \ecossurfspace from \unit[57]{GeV} to \unit[134]{GeV}. This work presents the first demonstration of an increase in the observed muon charge ratio between functionally equivalent shallow and deep underground detectors.  The most likely source of this increase is the greater probability that a muon in the deep detector data sample results from kaon production in extensive air showers.  A fit has been performed to a simple parametric model of muon production from pion and kaon parents, and the results of the fit support this interpretation of the combined MINOS Near and Far Detector charge ratio measurements.\\

\section{Acknowledgments}
This work was supported by the US DOE, the UK STFC, the US NSF, the State and University of Minnesota, the University of Athens, Greece and Brazil's FAPESP and CNPq. We are grateful to the Minnesota Department of Natural Resources, the crew of Soudan Underground Laboratory, and the staff of Fermilab for their contributions to this effort. 

\appendix*

%\begin{acknowledgments}
%\end{acknowledgments}

\bibliography{chargeratio}

\end{document}